\documentclass[journal]{IEEEtran}

\usepackage [english]{babel}
\usepackage [autostyle, english = american]{csquotes}
\MakeOuterQuote{"}
\IEEEoverridecommandlockouts

\usepackage[pdftex]{graphicx}
\usepackage{epstopdf}
\usepackage{amsmath}
\usepackage{amsfonts}
\usepackage{balance}

\usepackage{cite}

\usepackage{hyperref}
\usepackage{amssymb}
\usepackage{bbm}

\usepackage{array}
\usepackage{multirow}
\usepackage{color}
\usepackage{tablefootnote}

\usepackage[prependcaption,colorinlistoftodos]{todonotes}
\usepackage[mathscr]{euscript}

\newtheorem{proof}{Proof}
\newtheorem{theorem}{Theorem}
\newtheorem{definition}{Definition}
\newtheorem{assumption}{Assumption}
\newtheorem{remark}{Remark}

\newcommand{\jw}{\textcolor{black}}
\newcommand{\jdw}{\textcolor{black}}
\newcommand{\jd}{\textcolor{black}}
\newcommand{\jjdw}{\textcolor{black}}
\newcommand{\jdjdw}{\textcolor{black}}
\newcommand{\jdwjdw}{\textcolor{black}}
\newcommand{\icl}{\textcolor{black}}
\newcommand{\il}{\textcolor{black}}
\newcommand{\ill}{\textcolor{black}}
\newcommand{\jjddww}{\textcolor{black}}
\newcommand{\icc}{\textcolor{black}}
\newcommand{\jwrev}{\textcolor{black}}
\newcommand{\iclr}{\textcolor{black}}
\newcommand{\ilr}{\textcolor{black}}
\newcommand{\jdwrev}{\textcolor{black}}
\newcommand{\ilrr}{\textcolor{black}}
\newcommand{\lir}{\textcolor{black}}

\begin{document}

\title{Control of AC-AC interlinking converters for multi-grids}

\author{Jeremy D.~Watson and Ioannis Lestas
\thanks{Jeremy Watson is with the Department of Engineering, University of Canterbury, New Zealand. Ioannis Lestas is with the Department of Engineering, University of Cambridge, 
Cambridge CB2 1PZ, United Kingdom. Emails: {jeremy.watson@canterbury.ac.nz, icl20@cam.ac.uk}. 
This work was supported by European Research Council (ERC) starting grant 679774.}}
\maketitle

\begin{abstract}

\jdwjdw{This paper considers} the control of AC-AC interlinking converters (ILCs) in a multi-grid network. We overview the control schemes in the literature and propose a passivity framework for the stabilization of multi-grid networks, considering both AC grid-following and AC grid-forming behavior for the ILC connections. We then analyze a range of AC/AC interlinking converter control methods derived from \jdw{the} literature and propose suitable controllers for this purpose including both AC grid-forming and grid-following behavior. The controller we propose is partially grid-forming; \jdw{in particular, it is based on a} combination of \jdw{a} grid-following and \jdw{a} grid-forming converter \jdw{to improve the} stability properties \jdw{of the network}. Simulation results and theoretical analysis confirm that the proposed ILC control designs are appropriate for the multi-grid network.

\end{abstract}



\IEEEpeerreviewmaketitle

\section{Introduction}

AC multi-grids, or segmented AC grids, are becoming increasingly common, and are relevant both for microgrids, as well as for larger-scale "grids of grids" \cite{bellmunt2021}. The segments, or sub-grids, are connected by power converters, either via high voltage DC (HVDC) or back-to-back AC/AC interlinking converters (ILC). \jdw{In this paper, we consider the case of multiple AC subgrids \jdwjdw{or microgrids} connected via AC-AC ILCs to form a larger multi-grid. We will use the term "microgrid" (MG) in this paper broadly, referring to subgrids that make up the multi-grid. The analysis is hence not limited to low voltage microgrids, and \jdwjdw{we} include the IEEE 39 bus system in our examples.}

While ILCs for interlinking AC and DC sub-grid in hybrid AC/DC grids have attracted some analysis \cite{loh2013,unamuno20152,luo2016,malik2017,malik2018,CDC_paper,yoo2019,TPS_paper,TCST_paper,soler2023}, considerably less attention has been paid to the control of ILCs that connect two AC grids. Two AC grids can be connected by a transformer, however, \jdwjdw{the advantages of using an ILC include} asynchronous connection \jdwrev{(\ilrr{of microgrids with possibly} different frequencies or voltages)} as well as active and reactive power flow control. \jdwrev{Moreover, as we will show in this paper, it is possible to use the AC-AC ILC to \ilrr{provide} grid-forming services to improve system strength.}

Usually, an AC-AC ILC is comprised of two back-to-back AC-DC converters. \jdwjdw{It is possible to connect two AC grids without an intermediate DC stage using a matrix converter, however, voltage coupling and control are difficult \cite{ordono2019}.} Therefore, in practice ILCs for AC grids generally use an intermediary DC stage formed by the DC sides of the two AC-DC converters. The DC stage of a back-to-back AC/AC ILC can also be used to connect to a DC grid, thus forming an AC/DC/AC energy hub. The AC/AC ILC is therefore a practically relevant topology, however, its control has not yet been analyzed in detail from the perspective of network stability and scalability. 

Although the literature is limited, a few control schemes have been suggested. Most are AC grid-following (GFL): \jdwrev{i.e.,} the AC-DC converter(s) synchronize to the AC grid usually via a phase-locked loop (PLL) and control their power transfers. \jdwjdw{Frequency and voltage support may be provided via adjusting power transfers in response to frequency and/or voltage changes.} Other schemes are AC grid-forming (GFM), where the ILC converters set their AC frequency and voltage and thus directly support the frequency and voltage regulation of the AC grid. The issue of grid-following vs grid-forming behaviors for AC/AC ILCs has significant implications for the stability of the network. We also consider how each AC sub-grid is able to support other sub-grids through the ILC, e.g. providing inertial support.

In this paper, we examine control strategies for an AC-AC ILC with a shared DC bus, focussing on control strategies that are able to autonomously provide frequency control or support to the AC grids. While helpful insights are provided by recent work examining the small and large-signal stability of autonomous AC microgrids interconnected by back-to-back converters \cite{naderi20201,naderi20202}, we focus on finding conditions for stability {\icl{ in more general network configurations. In particular,} we derive conditions for a multi-grid network regardless of topology. Our conditions are also decentralized, \jdwrev{i.e.,} stability can be deduced from a local condition at each ILC. We propose for the first time (to our knowledge) a passivity framework for the stabilization of multi-grid networks, and consider a wide range of possible AC/AC ILC control schemes.

The contributions of this paper may therefore be summarized as follows:
\begin{enumerate}
    \item We review the literature regarding the control of AC/AC ILCs, classifying and analyzing possible schemes for outer control loop design. 
    \item We propose a passivity framework for multi-grid networks.  Focusing especially on the ILC, we find a decentralized condition \icl{through} which stability may be guaranteed. We also consider the case where both AC grid-following and grid-forming ILCs are present.
    \item We propose a controller for multi-grid ILCs which achieves our control objectives, using the concept of partially grid-forming AC/AC ILCs. We illustrate the effectiveness of our approach via simulations.
\end{enumerate}

\section{Background and literature review}\label{lit}
We start by reviewing the literature and classifying control schemes in Table \ref{tab:litacac}, focusing on the outer control loop design which is crucial for frequency regulation, stability, and power sharing. \jdwjdw{\jdwrev{We thus focus on schemes in which the ILC autonomously regulates or at least supports} the AC grid frequencies. Schemes using a central operator to send power commands (e.g. \cite{majumder2014,naderi20201,naderi20202}) fall outside the scope of this paper.} 

\begin{table}[ht]
	\centering
	\caption{\jwrev{Overview of control strategies for AC/AC interlinking converters.}}
		\begin{tabular}{c|c|c|c|c}
			 & Scheme & $V_{dc}$ regulation & GFM & Info. Req. 	\\
		    \hline
                \cite{tang2015} & Freq. droop & One VSC & No & None \\
		    \cite{bala2009} & Dual Freq. droop & Shared\tablefootnote{via PI control affecting the power setpoints of both VSCs.} & Yes & None \\
		    \cite{nutkani2013} & Dual Freq. droop & None\tablefootnote{The idea is to balance the power of the two VSCs to avoid disturbing the DC voltage. However, in practice small losses in the converter require a small term to regulate the DC voltage as in the authors' later work \cite{nutkani2015}.} & No & None \\
		    \cite{susanto2014} & Dual AC/DC droop & Shared\tablefootnote{via DC droop.} & No & None \\
		    \cite{nutkani2015} & Dual Freq. droop\tablefootnote{with deadband.} & Shared\tablefootnote{via PI controller affecting the power setpoints of both VSCs.} & No & None \\
                \cite{an2022} & Dual Freq. droop\tablefootnote{(2) in \cite{an2022}; other controllers are also suggested.} & One VSC & No & None \\
		    \cite{sun2015} & Freq. droop\tablefootnote{This paper considers the slightly different scenario of the connection of 3 single-phase AC microgrids with each other and a 3-phase AC microgrid.} & One VSC & Partial & None \\
		
		    \cite{goyal2016} & Angle droop & One VSC & Partial & Yes\tablefootnote{Droop coefficients of the DGs are communicated.} \\
            \cite{yoo2017} & Dual Freq. droop & One VSC & No & None \\
            \cite{mazidi2020} & Dual Freq. droop & One VSC & No & Yes\tablefootnote{Although communication is used for several aspects of this controller, it is not required for the dual frequency-droop operation of the ILC.} \\
            \cite{pham2021} & Dual Freq. droop\tablefootnote{A VSG-based control, which is similar to droop, is also presented.} & One VSC & No & None \\

		\end{tabular}
	\label{tab:litacac}
\end{table}

\jdwrev{Little comparative analysis of these has been performed in \ilrr{the} literature to-date, a gap which this paper intends to address.} In addition to the schemes in Table \ref{tab:litacac}, \jdw{controllers proposed for AC/DC interlinking conversion may be relevant since the AC/AC ILC is effectively two AC/DC ILCs connected back-to-back}. \jdwrev{Since power-frequency droop is ubiquitous, we require the ILC to equalize the normalized frequencies of the two AC subgrids it connects (see Section \ref{ctrlobj}) and thus achieve power-sharing.} We therefore focus exclusively on controllers which achieve this in Table \ref{tab:litacdc}. Thus, we do not consider some \icl{other} 
interesting schemes for hybrid AC/DC network control, such as the hybrid angle control in \cite{hac2022}.
\begin{table}[ht]
	\centering
	\caption{AC/DC ILC controllers also suitable for AC/AC ILCs.}
		\begin{tabular}{c|c|c}
			Reference & Scheme description & Grid-forming	\\
		    \hline	
		    \cite{jouini2016, arghir2018, CDC_paper,TCST_paper} & Matching control & Yes \\
		    \cite{loh2013,yoo2019,TPS_paper} & (Integral) Dual AC/DC Droop & No \\
		\end{tabular}
	\label{tab:litacdc}
\end{table}

In the rest of this paper, we will analyze \jdwrev{and compare these schemes. In particular, the dual frequency droop schemes \cite{bala2009, nutkani2013, nutkani2015, an2022, yoo2017, mazidi2020, pham2021} are summarized and presented in Section \ref{sect:dfd}; dual AC/DC droop \cite{susanto2014, loh2013,yoo2019,TPS_paper} in Section \ref{sect:dacd}; matching control \cite{jouini2016, arghir2018, CDC_paper,TCST_paper} in Section \ref{sect:dmc}; frequency droop schemes \cite{tang2015, sun2015} (with a little modification as discussed later) in Section \ref{sect:gffd}; we also notice that dual AC/DC droop can be re-written as a grid-forming controller and analyze this in Section \ref{sect:gfdd}. Angle droop is not analyzed as this is unable to regulate MG frequency without extra communication requirements. We consider} their suitability for a decentralized stability guarantee using passivity. We discuss considerations regarding grid-forming and grid-following behavior and inertial support. \jw{We then} propose a new solution and illustrate our findings with our case studies.

\section{Preliminaries and notation}\label{prelim}

The multi-grid model is defined by a connected graph $(N,Z)$, where $N := \{1, 2, ..., |N|\}$ is the set of microgrids (MGs), and $Z \subset N \times N$ is the set of interconnecting ILCs. Each microgrid (MG) $i \in N$ is therefore connected to at least one ILC $z \in Z$. \jw{The ordering of the MGs in $N$ and the ILCs in $Z$ is arbitrary. We also enumerate the ILCs as follows: an ILC $z \in Z$ also has a number $l \in \{1, 2, ..., |Z|\}$ and this number corresponds to the ordering in $Z$.} Each ILC $z \in Z$ has two connections that connect to two MGs, and these connections are enumerated by $Q$ where the connections are ordered in the same way as the ILCs in $Z$, \jdwrev{i.e.,} the first two entries in $Q$ represent the connections of the first ILC in $Z$, followed by the connections of the second ILC, etc. 
\jw{Hence $|Q| = 2|Z|$.}



%

\section{Passivity conditions for multi-grids}\label{stab}

\subsection{Passivity and related definitions}
One promising tool to ensure stability in a scalable way is passivity theory. Passivity-based approaches have been proposed when AC networks (e.g. \cite{harnefors2016,kasis2017}) or DC networks (e.g. \cite{gu2015,soloperto2018,cucuzella2019}) are individually studied, as well as for hybrid AC/DC networks \cite{TPS_paper}.

\jdwrev{The idea behind passivity-based analysis of power systems is to note that power systems dynamics can generally be written into a form where there are two subsystems connected via negative feedback, typically corresponding to buses and lines, \ilrr{respectively}. If it is possible to passivate both (multi-input multi-output) subsystems, \ilrr{\lir{stability} of the network can be deduced from a main result in passivity theory which states} 
that the negative feedback interconnection of two passive systems is stable and passive \cite{khalil1991}.} Since the power line equations have natural passivity properties, \jd{passivation of the bus dynamics is sufficient for stability}.




\jdwjdw{However, it is difficult to ensure the passivity of all buses \cite{harnefors2016}, especially since the most natural choice of input and output variables ($DQ$ voltages and currents in a common reference frame) leads to a passivity criterion that is impossible to satisfy with common frequency control schemes \cite{dey2021}.}

\jdwjdw{This study is instead focused on the design of the controller of the ILCs between AC networks, \jdwrev{i.e.,} the stabilization of individual MGs is beyond our scope.} We will therefore assume that individual MGs are stable and satisfy a passivity property with regards to input power and output frequency, a condition which is generally fulfilled by \jdjdw{frequency-droop} controlled MGs and is obviously less restrictive than requiring every bus within every MG to be passive.

We start by defining the passivity property \cite{khalil1991, TPS_paper}. Consider a dynamical system with $m$ inputs, $n$ states, and $m$ outputs of the form:
\begin{subequations} \label{eq:ss}
\begin{align}
    \dot{x} &= f(x,u)\\
    y &= g(x,u)
\end{align}
\end{subequations}
where $u(t) \in \mathbb{R}^m$ is the input vector, $x(t) \in \mathbb{R}^n$ the vector of states, and $y(t) \in \mathbb{R}^m$ the output, $f$ is locally Lipschitz, and $g$ is continuous.

\begin{definition}[Passivity \cite{khalil1991,TPS_paper}]\label{def:isp}
The system \eqref{eq:ss} is locally passive about an equilibrium point with constant input $\bar{u}$ and constant state $\bar{x}$ if there exists open neighbourhoods $U$ of $\bar{u}$ and $X$ of $\bar{x}$ and a continuously differentiable function $W(x)$ \ill{(known as the storage function)} such that
\begin{gather}\label{eq:storage}
\dot{W}(x) \leq (u-\bar{u})^T(y-\bar{y}) .
\end{gather}
for all $x\in X$ and $u\in U$. \jd{The system is locally output strictly passive if the definition above holds with the right hand side of \eqref{eq:storage} replaced by $(u-\bar{u})^T(y-\bar{y}) -\phi(y-\bar{y})$ where $\phi$ is a positive definite function.}
\end{definition}

\begin{assumption}
    \jjdw{For systems of the form \eqref{eq:ss} which are locally passive about an equilibrium point $x^*$, we will assume that the storage \ill{function $W(x)$ 
    has} a strict local minimum at $x^*$. }\end{assumption}

\icl{This is a relatively mild assumption which is always satisfied for linear systems when these are observable and controllable~\cite{khalil1991}.}

\begin{remark}
If the system \eqref{eq:ss} is linear and the passivity property holds locally about an equilibrium point, it also holds globally, \jdwrev{i.e.,} for any other equilibrium point and for any deviation from an equilibrium point. Therefore for linear systems which satisfy Definition \ref{def:isp}, we will refer to them as being "passive" (or "strictly passive") without making reference to a particular equilibrium point.
\end{remark}

We will also define two properties that will be useful for applying LaSalle's theorem to prove asymptotic stability, namely \jwrev{equilibrium-state observability and input observability. These technical definitions are in Appendix B in order to preserve the flow and readability of the paper.}

\subsection{Microgrid dynamics}
\jwrev{We consider the active power and AC frequency dynamics for each microgrid. This is therefore a decoupled analysis and AC voltage dynamics / reactive power are ignored. AC voltage dynamics are modelled in our second case study for verification purposes, but in general this is beyond the scope of our paper and is left to future work. }

\jdw{We consider a single-input single-output state-space representation of each MG $j \in N$,  with state vector $x_j$, of the form \eqref{eq:ss}. The input is the aggregate power entering the microgrid $u_j = p_j$ and the output is the microgrid frequency $y_j = \omega_j$. We are therefore assuming that all MGs have a single point at which all the ILC connections of that MG are located. We will discuss how this assumption can be relaxed in Section \ref{generalcase}. We are then ready to state the following assumption:}
\begin{assumption}\label{mgpass}
\jdw{Consider a state space representation of the dynamics of each MG $j \in N$ \jd{as in \eqref{eq:ss}}, with state vector $x_j$, input (power) $p_j$ and output (frequency) $\omega_j$. The system is}
\begin{enumerate}
\renewcommand{\theenumi}{\alph{enumi}}
    \item \jd{output strictly passive} \jw{about an equilibrium point $(p^\ast_j, x^\ast_j)$}, \jdwrev{i.e.,} the condition in Definition \ref{def:isp} is satisfied with input $p_j$ and output $\omega_j$.
    \item \jd{asymptotically stable about an equilibrium point $(p^\ast_j, x^\ast_j)$, \jdwrev{i.e.,} $x^\ast$ is an asymptotically stable equilibrium point for constant input $u^\ast$.}
    \item \jjdw{input observable about $(p^\ast_j, x^\ast_j)$ (Definition \ref{def:iobsrv})}.
\end{enumerate}
\end{assumption}
\begin{remark}
Note that \jwrev{in Assumption 2a)} we require each MG to have aggregate passive dynamics \jd{with respect to input power and output frequency}. This is less restrictive than requiring all buses within each MG to be passive, e.g. as in previous work \cite{TCST_paper}. \jwrev{This assumption is reasonable if the MG frequency response is dominated by the droop characteristic of one or more generators, which typically results in \ilr{dynamics 
from power to frequency dominated by a first order system} (as in \cite{chandorkar1993} for inverters and many further papers). \ilr{Higher order dynamics with sufficiently high damping can also be passive \lir{\cite{kasis2017}}. 
There are, however, cases where the MG dynamics may not be passive}
and this aspect should be verified in practice. Assumption 2b) is reasonable given \lir{that a MG} should be capable of running stably in an islanded or stand-alone setting.}
\end{remark}

\begin{remark}
 \jd{The condition \icl{of} input observability \jwrev{(Assumption 2c)} is satisfied for practical MG dynamics, \jdwrev{i.e.,} it states that any deviation in input power will be reflected on the (output) frequency of the MG.}
\end{remark}

\subsection{ILC dynamics}
\jdw{\icl{Each} ILC
\icl{with associated number $l$ has two connection points denoted as  $\rho$ and $\sigma$ respectively, and} is \icl{similarly}  modelled by a state-space representation \icl{as in \eqref{eq:ss}} with state vector $z_l$ and input $u_l = [\omega^x_{\rho}\mbox{ }\omega^x_\sigma]^T$ and output $y_l = [-p^x_{\rho}\mbox{ }-p^x_\sigma]^T$. \icl{In particular, the} inputs are the frequencies at the two connections of the ILC,
and the outputs are the powers entering the ILC from its connections (\jdwrev{i.e.,} the negative of the output power at each ILC connection), \icl{as illustrated in \il{Fig. \ref{fig:ilcnot}. \jwrev{These input/output variables imply grid-following ILCs}\footnote{When considering GFM ILCs, we will use the same notation except for changing the subscript to a capital $X$, \jd{in order} to distinguish between GFL and GFM ILCs.}}.}}
\begin{assumption}\label{ilcpass}
The dynamics of \icl{each ILC $l$ with state vector $z_l$ are:}
\begin{enumerate}
\renewcommand{\theenumi}{\alph{enumi}}
\item \jd{passive} \jw{about an equilibrium point $(u^\ast_l, z^\ast_l)$, \jdwrev{i.e.,}} the condition in Definition \ref{def:isp} is satisfied with input $u_l = [\omega^x_{\rho}\mbox{ }\omega^x_\sigma]^T$ and output $y_l = [-p^x_{\rho}\mbox{ }-p^x_\sigma]^T$.
\item \jjdw{equilibrium-state observable about $(u^\ast_l, z^\ast_l)$ (\il{Definition~\ref{def:eso}}).}
\end{enumerate}
\end{assumption}
\begin{remark}
\jwrev{In the rest of this paper, we will consider ILC control schemes and investigate whether they can satisfy this assumption. Assumption 3b) is generally satisfied for all the control schemes we consider (we verify this for our proposed controller in Appendix C) and is a technical point which we use to be able to prove asymptotic stability of the multi-MG network. We will show that Assumption 3 can be satisfied by reasonable control designs and present examples in the next section and in Section \ref{casestudy}. }
\end{remark}

\vspace{-2mm}
\begin{figure}[ht]
    \centering
    \includegraphics[width=0.49\textwidth]{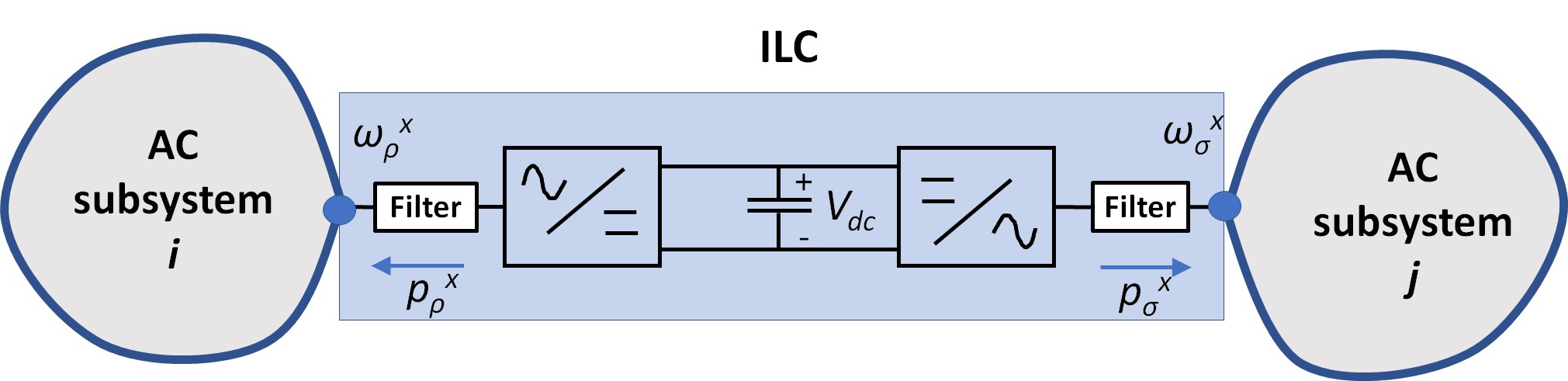}
    \caption{ILC notation.}
    \label{fig:ilcnot}
\end{figure}
\vspace{-5mm}
\subsection{Passivity conditions for AC-AC ILCs}
Having defined both the MG and ILC subsystems and presented passivity assumptions, we now illustrate the passivity framework (Fig. \ref{fig:ACAC}) for the simple example of two ILCs connecting three AC MGs (Fig. \ref{fig:ACACdiag}).
\begin{figure}[ht]
    \centering
    \includegraphics[width=0.45\textwidth]{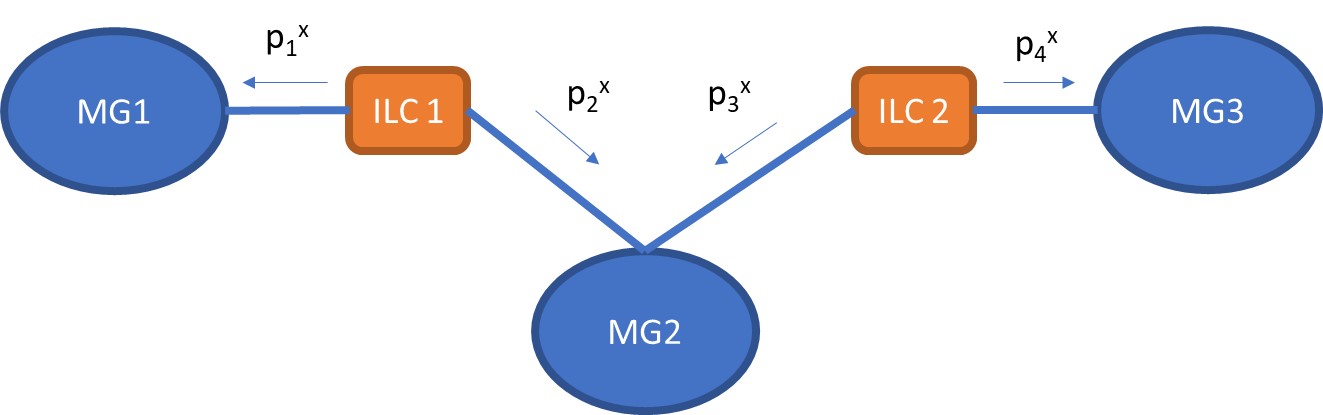}
    \caption{Network diagram.}
    \label{fig:ACACdiag}
\end{figure}

\begin{figure}[ht]
    \centering
    \includegraphics[width=0.45\textwidth]{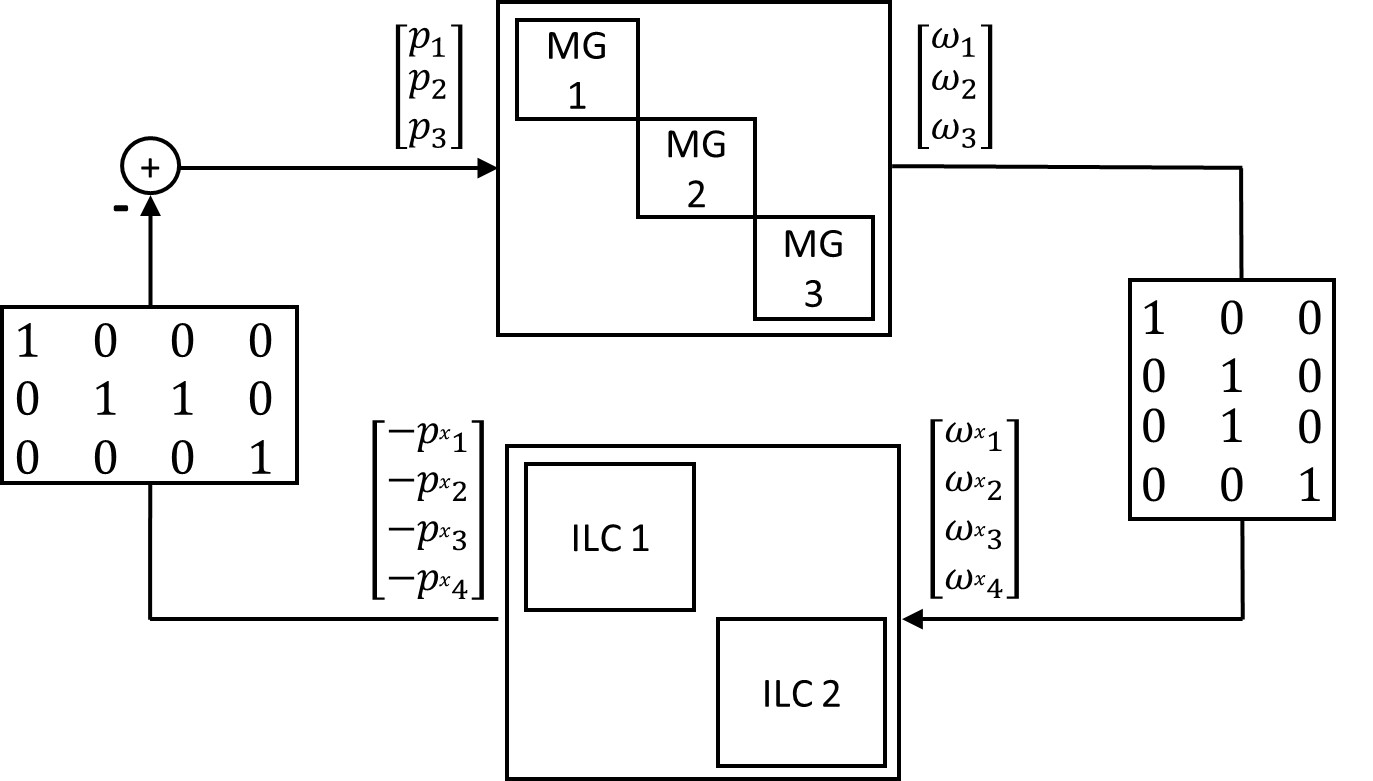}
    \caption{Representation of a network containing three AC MGs and two ILCs.}
    \label{fig:ACAC}
\end{figure}

Under the assumption that individual MGs satisfy a (strict) passivity property with regards to input power and output frequency (Assumption \ref{mgpass}), it follows that if both the ILCs are passive with respect to input frequency and (negative) output power, the multi-MG network will be stable. For larger networks of AC MGs, the same criterion is sufficient for stability. It is again possible to write the power network dynamics as the negative feedback interconnection of the MG dynamics and the ILC dynamics, as in Fig. \ref{fig:ACAC1},
\begin{figure}[ht]
    \centering
    \includegraphics[width=0.45\textwidth]{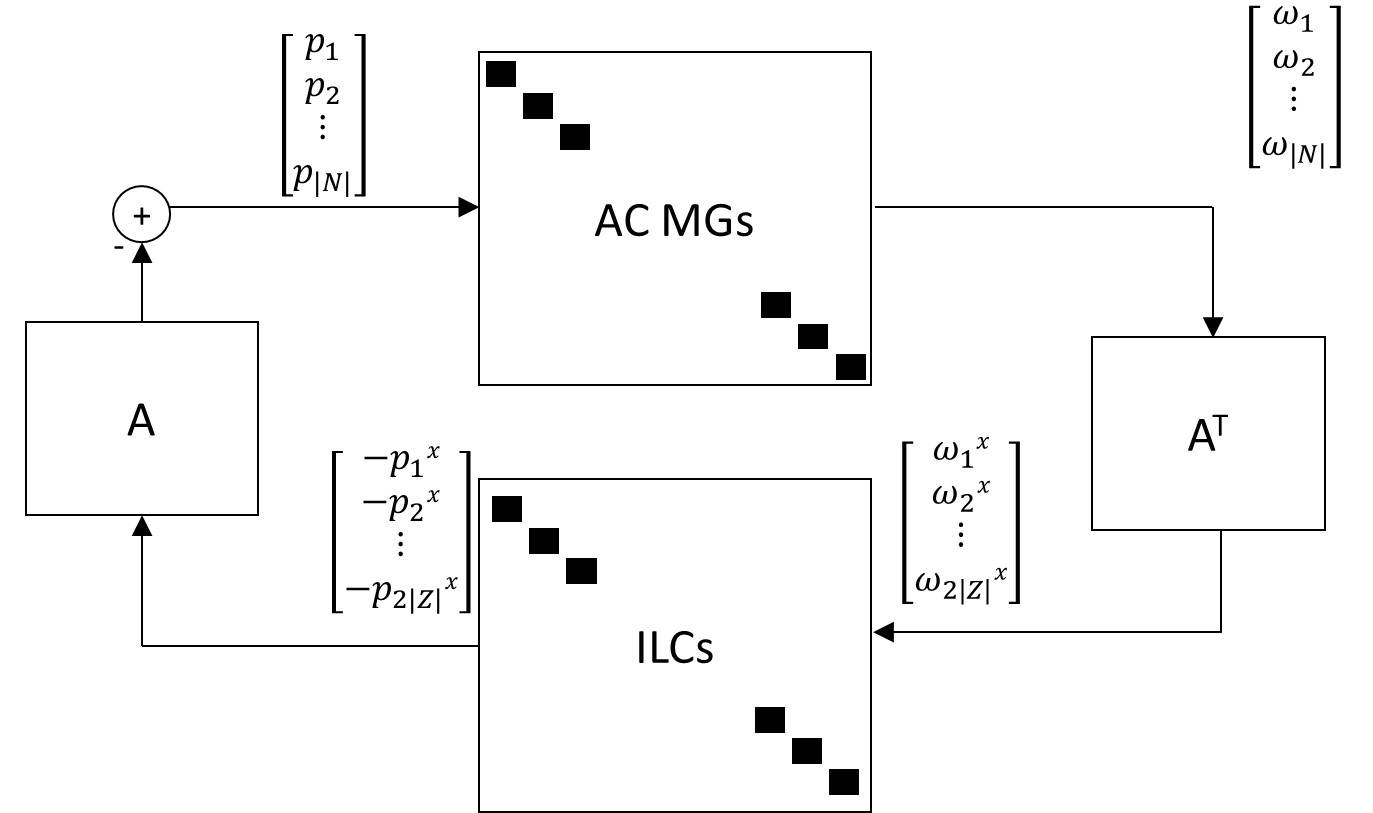}
    \caption{Representation of a general network with multiple MGs and ILCs, where each MG has a single connection point between it and the ILC(s).}
    \label{fig:ACAC1}
\end{figure}
where the matrix $A$ (with dimension $|N| \times |Q|$) has the following definition, using the enumeration of ILC connections defined earlier:
\begin{equation}\label{eq:ILCincidence}
    A_{i,\rho} =
    \begin{cases}
    1 &\text{if ILC connection ${\rho}$ is at MG $i$}\\
    0 & \text{otherwise}
    \end{cases}
\end{equation}


\jdwjdw{We now present} the stability result for multi-grid networks under the passivity assumptions (Assumptions \ref{mgpass} and \ref{ilcpass}).

\begin{theorem}\label{stabtheorem}
Suppose there exists an equilibrium $x^\ast = [x^\ast_1, x^\ast_2, ... x^\ast_{|N|}, z^\ast_1, z^\ast_2, ... z^\ast_{|Z|}]^T$ for the interconnected MG and ILC dynamics, for which Assumptions \ref{mgpass} and \ref{ilcpass} are satisfied. Then, such an equilibrium point is locally asymptotically stable.
\end{theorem}

\emph{Proof:} See Appendix A.

Theorem \ref{stabtheorem} follows from the fact that the multi-MG model is the negative feedback interconnection of \jd{a strictly passive system and passive system}. With some of the controllers in Section \ref{ctrl}, Assumption \ref{ilcpass} can be satisfied, and thus the stability of the network \jd{can be} guaranteed.

\subsection{General passivity conditions for ILCs}\label{generalcase}
There are two cases that deserve further attention. Firstly, the case that any MG has multiple ILC connections at different points within that MG. In this case, the framework must include the lines within each MG, which necessitates additional assumptions.
Secondly, the case where AC grid-forming ILCs are present. AC GFM ILCs take power as input, and the output is the AC frequency. For this reason, a different decomposition must be used.


The passivity-based stability criterion is still relevant. The system can be written as a negative feedback interconnection, with the grid-following ILCs and distribution/transmission lines forming one subsystem, while all the buses in each MG and the GFM ILCs
form the other. This contrasts with previous work on the passivity of ILCs \cite{TPS_paper}, where the ILCs were grouped with the bus dynamics. Grouping the GFL ILC dynamics together with that of the lines seems logical (as they both transfer power between buses) and provides a more straightforward extension to the multi-MG case.

For this case, it is helpful to define a directed\footnote{The choice of direction is arbitrary.} graph ($\mathcal{V}, \mathcal{E}$) where $\mathcal{V} := {1, 2, ... |\mathcal{V}|}$ is the set of nodes representing all buses in all MGs and all GFM ILC connections, and $\mathcal{E} \subset \mathcal{V} \times \mathcal{V}$ is the set of lines \jd{and GFL ILCs. The GFL ILCs are represented by $\Lambda \subset \mathcal{E} \subset \mathcal{V} \times \mathcal{V}$}, \jdwrev{i.e.,} an edge $(i,j) \in \Lambda$ \jd{is associated with} a GFL ILC which connects node $i$ to node $j$. \jd{These nodes which are connected via a GFL ILC $z \in \Lambda$ are also called "GFL ILC connections" and these connections are enumerated by $L := \{1, 2, ..., |L|\}$. As} before the connections are ordered in the same way as the ILCs in $\Lambda$, and $|L| = 2|\Lambda|$. We also define the following constants: $m$ - the number of lines (number of edges in $\mathcal{E}$ minus the number of edges in $\Lambda$); $\nu$ - the number of MG buses in the multi-MG network; $\mu$ - the number of GFM ILC connections (twice the number of GFM ILCs); and $\lambda = |L|$ - the number of GFL connections. \jd{It is important to note that GFM ILC \icl{connection points} are nodes, whereas GFL ILCs are represented by edges.}

\jdw{Given these definitions, the passivity framework \jdw{under the new decomposition} is illustrated in Fig. \ref{fig:ACDC2}, where the matrix $\mathcal{M}$ is the incidence matrix of the graph ($\mathcal{V}, \mathcal{E}$), and the matrix $\mathcal{A}$ is an incidence-like matrix defined similarly to \eqref{eq:ILCincidence}:}
\begin{equation}\label{eq:ILCincidenceM}
    \mathcal{M}_{i,\rho} =
    \begin{cases}
    1 &\text{if $i$ is the source of edge $\rho = (i,:)$ }\\
    -1 &\text{if $i$ is the sink of edge $\rho = (:,i)$ }\\
    0 & \text{otherwise}
    \end{cases}
\end{equation}
\begin{equation}\label{eq:ILCincidenceA}
    \mathcal{A}_{i,\rho} =
    \begin{cases}
    1 & \text{if GFL ILC connection ${\rho}$ is at node $i$}\\
    0 & \text{otherwise}
    \end{cases}
\end{equation}
\begin{figure}[ht]
    \centering
    \includegraphics[width=0.45\textwidth]{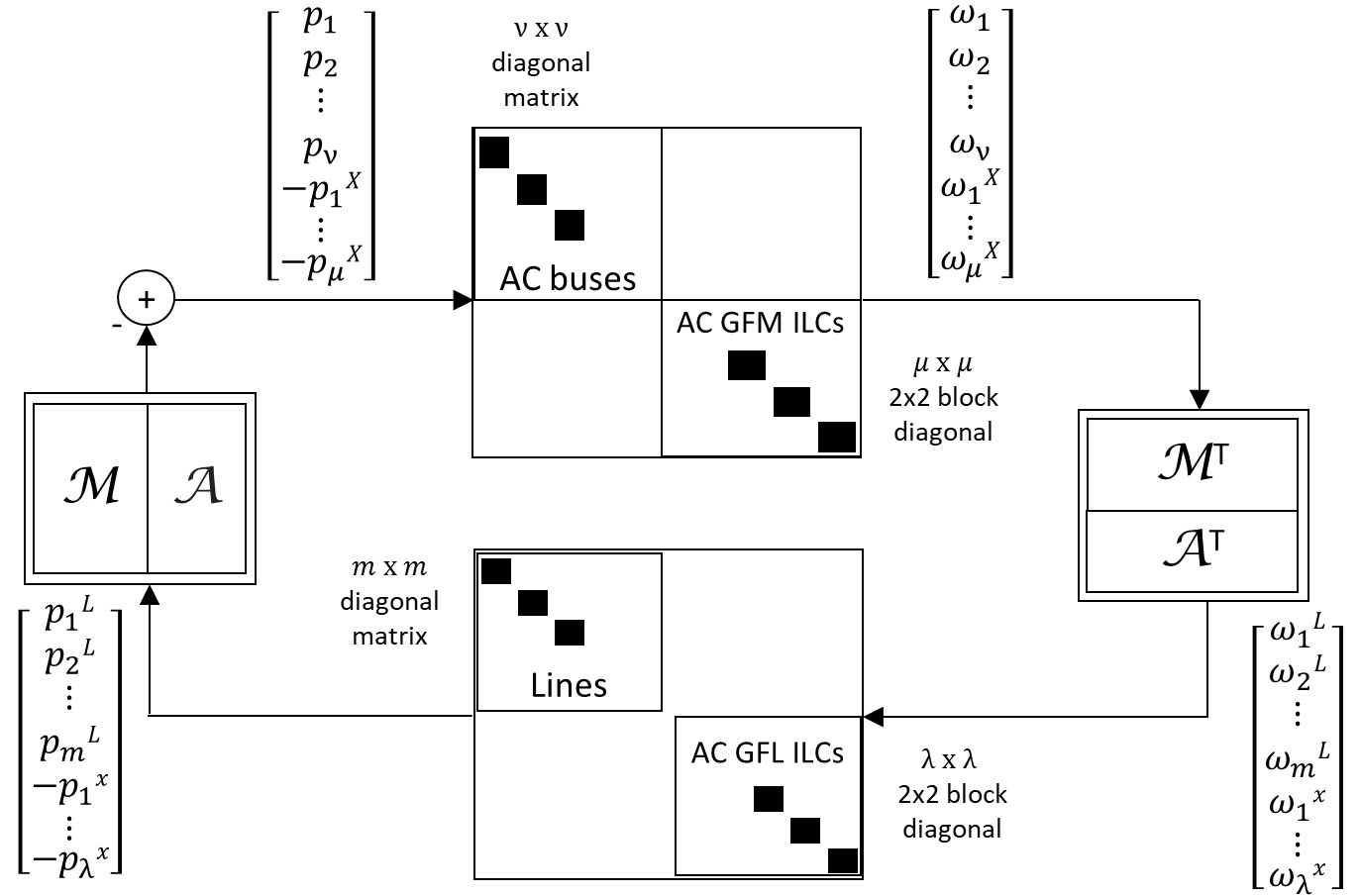}
    \caption{Representation of a larger network with multiple MGs and ILCs, where any MG may have multiple connection points to various ILCs.}
    \label{fig:ACDC2}
\end{figure}
However, additional assumptions are required to derive sufficient decentralized stability conditions for the ILCs. 
These assumptions (1-2 below) together with the ILC conditions for both GFL and GFM behavior (3-4 below) are:

\begin{enumerate}
    \item The dynamical nature of the lines and their conductances are neglected (similarly to \cite{kasis2017}).
    \item All MG buses $i \in \mathcal{V}$ are passive with input $p_i$ and output $\omega_i$.
    \item Assumption \ref{ilcpass} holds for the GFL ILCs $z \in \Lambda$.
    \item Each GFM ILC with associated connections $i$, $j$ is passive \jd{about an equilibrium point}, \jdwrev{i.e.,} the condition in Definition \ref{def:isp} for passivity is satisfied with input $[-p^x_{i}\mbox{ }-p^x_{j}]^T$ and output $[\omega^x_i\mbox{ }\omega^x_j]^T$.
\end{enumerate}
Stability would then follow trivially from the interconnection of two passive subsystems. Hence, for the GFM control designs in section \ref{ctrl}, we will consider their passivity properties with input $[-p^x_{i}\mbox{ }-p^x_{j}]^T$ and output $[\omega^x_i\mbox{ }\omega^x_j]^T$. 

\jwrev{However, it should be noted that the additional assumptions may in some cases be \ilr{more difficult to satisfy}
e.g. in medium or especially low-voltage grids, the line conductance cannot generally be neglected; furthermore, the requirement is now that all MG buses be passive as opposed to merely the overall power-frequency dynamics of each MG.}




\section{Control design for AC-AC ILCs}\label{ctrl}

\subsection{Control objectives}\label{ctrlobj}
Achieving autonomous power sharing between the AC grids requires their per-unit frequencies to be equalized (assuming droop-controlled sources in each grid, as is ubiquitous) \cite{yoo2017}. This is analogous to the case in hybrid AC/DC networks, where normalizing the AC frequency and DC voltage enables power-sharing \cite{TCST_paper,TPS_paper}. Our control objectives are therefore:
\begin{enumerate}
    \item \emph{Power sharing:} The ILC should control the powers and/or frequencies of AC subsystems such that frequency droop-controlled AC sources contribute in proportion to their droop coefficients.
    \item \emph{Stability:} Stability guarantees for the network via local conditions on the MG and ILC dynamics.
\end{enumerate}

The steady-state normalization of AC frequencies to achieve the power-sharing objective can be achieved by many of the schemes in Tables \ref{tab:litacac} and \ref{tab:litacdc} (with the straightforward addition of an integral term in one case). \jdw{We will present and then analyze these schemes, starting with the reduced-order model of the converter itself, and then proceeding to the design of the outer control loops.} 

\subsection{ILC models}
\jdw{In this section, we present the ILC models which we will use for theoretical analysis. \jwrev{Because our study focuses on the control design of the slower outer loop dynamics (discussed in the next sections) and their interactions with microgrid dynamics, we use a reduced-order model of the ILC \cite{TPS_paper}. As such, we take frequency as the input and power as an output, ignoring fast dynamics such as inner loops, converter LC (or LCL) filters, etc.}\footnote{\jwrev{However, we will verify our proposed controller on a detailed simulation model in Section \ref{casestudy}.}} We present three models, depending on whether the ILC is AC grid-following, AC grid-forming, or partially grid-forming.}

\jdw{It is crucial to note that $P_{ref,1}$, $P_{ref,2}$, $\omega_{ref,1}$, and $\omega_{ref,2}$ in \eqref{eq:gfl_ilc}-\eqref{eq:mix_ilc} are not constants, but are instead variables that are determined dynamically by the outer control loops. These outer control loops \jd{which are described in the following sections} include feedback and allow the ILC to synchronize to the MGs it connects. When the outer control loop equations are taken in conjunction with the appropriate ILC model (\eqref{eq:gfl_ilc} for GFL ILCs, \eqref{eq:gfm_ilc} for GFM ILCs, etc.) this creates the controlled ILC model we analyze. \jdw{We will use the indices 1 and 2 as the ILC connections for simplicity of notation.}}

\subsubsection{Grid-following ILC model}
\jdw{We first consider grid-following ILCs, which measure the input frequencies $\omega^x_1$ and $\omega^x_2$ (not directly shown in \eqref{eq:gfl_ilc} as embedded in the control via $P_{ref,1}$ and $P_{ref,2}$) at their connections and control their output power $p^x_1$, $p^x_2$.} The model is:
\begin{subequations}\label{eq:gfl_ilc}
\begin{align}
    \tau_1 \dot p^x_1 & = - p^x_1 + P_{ref,1} \\
    \tau_2 \dot p^x_2 & = - p^x_2 + P_{ref,2} \\
    C \dot{\tilde{V}}_{dc} & = - \frac{p^x_1}{\tilde{V}_{dc} + V_{dc,ref}} - \frac{p^x_2}{\tilde{V}_{dc} + V_{dc,ref}} - K_{dc}\tilde{V}_{dc}
\end{align}
\end{subequations}
where the subscripts $1$ and $2$ denote the two connections of the AC-AC ILC, $\tau_1$, $\tau_2$, are the converters' time constants respectively, $p^x_1$, $p^x_2$ are deviations from the nominal in the DC to AC power (\jdwrev{i.e.,} power leaving the ILC) of each inverter, $P_{ref,1}$, $P_{ref,2}$ are the reference values set by the outer control loop, $C$ is the DC capacitance, $K_{dc}$ is a coefficient representing DC support or load, $\tilde{V}_{dc}$ is the DC voltage deviation from the nominal, $V_{dc,ref}$ the nominal DC voltage, and $i_{dc}$ is the DC input current.

\jwrev{Grid-following converters require PLLs, and in our model the PLL is encapsulated in the first-order \iclr{lag $\tau_i$} for each grid-following converter to simplify the model for analysis. We \ilr{have also} 
investigated the behaviour of typical higher-order PLLs in both the time and frequency domains to verify that they behave similarly (in a small-signal sense) to our first-order model in the time-scale we are analyzing. 
PLLs may also face large-signal synchronization issues when the grid is weak, however, this is beyond the scope of our paper.}

\subsubsection{Grid-forming ILC model}
The back-to-back AC/AC ILC can set the AC frequency at its AC connections, \jdwrev{i.e.,} AC grid-forming operation. \jdw{In this case, the input to the ILC is powers $p^x_1$ and $p^x_2$, and the output is the frequencies $\omega^x_1$ and $\omega^x_2$ which are set by the ILC.} The model is:
\begin{subequations}\label{eq:gfm_ilc}
\begin{align}
    \omega^x_1 & = \omega_{ref,1} \\
    \omega^x_2 & = \omega_{ref,2} \\
    C \dot{\tilde{V}}_{dc} & = - \frac{p^x_1}{\tilde{V}_{dc} + V_{dc,ref}} - \frac{p^x_2}{\tilde{V}_{dc} + V_{dc,ref}} - K_{dc}\tilde{V}_{dc}
\end{align}
\end{subequations}
where $\omega^x_1$, $\omega^x_2$ are frequency deviations from the nominal, and other variables and parameters are as in \eqref{eq:gfl_ilc}. \jd{Since the converter is able to set the frequency \jjdw{very quickly}, we neglect $\tau_1$ and $\tau_2$ in this model (consistent with the fact that our analytical results focus only on the slower dynamics of the outer control loops).} 

\subsubsection{Partially grid-forming ILC model}\label{sectpartiallygfm}
We will also consider the case where one VSC is AC grid-forming, while the other is AC grid-following. In this case, the model is:
\begin{subequations}\label{eq:mix_ilc}
\begin{align}
    \tau_2 \dot p^x_2 & = - p^x_2 + P_{ref,2} \\
    \dot{\eta} &= \omega_{ref,1} - \omega_{1} \label{eq:ind1} \\
    p^x_1 &= B \sin \eta \label{eq:ind2}\\
    C \dot{\tilde{V}}_{dc} & = - \frac{p^x_1}{\tilde{V}_{dc} + V_{dc,ref}} - \frac{p^x_2}{\tilde{V}_{dc} + V_{dc,ref}} - K_{dc}\tilde{V}_{dc}
\end{align}
\end{subequations}
where $\omega_{1}$ is the grid frequency \jdw{deviation at the GFM side} of the VSC, $\eta$ is the angle between the first VSC and the grid, and $B > 0$ is a constant based on the susceptance of the inductive filter. \jdw{\eqref{eq:ind1}-\eqref{eq:ind2} model the inductor filter \jd{via $B$} which relates the frequency deviation of the GFM side to the power output from this side of the ILC into the connected MG. Since the GFL side takes frequency as an input and power as an output, \jd{we reformulate the GFM side to have \il{an} \icl{analogous} 
input and output \icl{when a passivity analysis is carried out}. Hence,} the model \eqref{eq:mix_ilc} \icl{is considered to have} inputs $\omega^x_1$ and $\omega^x_2$ and outputs $p^x_1$, $p^x_2$ similarly to a GFL ILC. }

\subsection{AC Grid-following controllers}
\jdw{In this case, the controller sets the reference powers $P_{ref,1}$ and $P_{ref,2}$ according to the outer loop design which we discuss in this section.
The equations which set the reference powers $P_{ref,1}$ and $P_{ref,2}$ are used in conjunction with the ILC model \eqref{eq:gfl_ilc} to form the reduced-order dynamical model of the AC/AC ILC. }

\subsubsection{Dual Frequency Droop}\label{sect:dfd}

In dual frequency droop, two frequency droop schemes are simultaneously used to control power transfer through the ILC:
\begin{equation}\label{eq:dfd_0}
    P_{ref,1} = -K_{\omega1}\omega_1 + K_{\omega2}\omega_2
\end{equation}
To equalize the frequencies, an integral controller can be used as in \cite{yoo2019} for example.
\begin{subequations}\label{eq:dfd_1}
\begin{align}
    \dot \xi &= -K_{\omega1}\omega_1 + K_{\omega2}\omega_2 \\
    P_{ref,1} &= -K_{\omega1}\omega_1 + K_{\omega2}\omega_2 + K_i \xi
\end{align}
\end{subequations}

If no load or generation is connected to the DC bus, in theory it is possible to simply set $P_{ref,2} = -P_{ref,1}$ and the DC voltage would remain constant as the input and output power would always be equal. However, losses in the converter and unmodelled discrepancies between the inverters necessitate some form of DC voltage regulation.

As seen in the literature survey, DC voltage control can be achieved either by one converter as in \cite{yoo2019}, or shared between both as in \cite{nutkani2015}. In the former case, one converter \jdw{(which we will designate as VSC 2)} focuses solely on DC voltage regulation, where a PI loop is typically used:
\begin{subequations}\label{eq:dfd_2}
\begin{align}
    \dot \zeta &= \tilde{V}_{dc} \\
    P_{ref,2} &= K_{pdc}\tilde{V}_{dc} + K_{idc} \zeta
\end{align}
\end{subequations}
We will call the combination of \eqref{eq:dfd_1} and \eqref{eq:dfd_2} \emph{Dual Frequency Droop (1)}.

In the latter case \jd{(\jdwrev{i.e.,} shared DC voltage regulation between the two converters)}, which we will call \emph{Dual Frequency Droop (2)}, an integral term ($P_{dc}$) is still calculated but \jd{it is} shared between both inverters:
\begin{subequations}\label{eq:dfd_shared}
\begin{align}
    \dot \xi &= -\omega_1 + \omega_2 \\
    \dot \zeta &= \tilde{V}_{dc} \\
    P_{dc} &= K_{pdc}\tilde{V}_{dc} + K_{idc} \zeta \\
    P_{ref,1} &= -K_{\omega1}\omega_1 + K_{\omega2}\omega_2 + K_i \xi + P_{dc} \\
    P_{ref,2} &= K_{\omega1}\omega_1 - K_{\omega2}\omega_2 - K_i \xi + P_{dc}
\end{align}
\end{subequations}
The equations for the dual frequency droop scheme are given by \eqref{eq:gfl_ilc},\eqref{eq:dfd_1},\eqref{eq:dfd_2} for the first case (DC regulation achieved by one inverter only), and \eqref{eq:gfl_ilc},\eqref{eq:dfd_shared} for the second case (shared DC regulation).

\subsubsection{Dual AC/DC Droop} \label{sect:dacd}

In this case, the AC/DC/AC ILC is simply treated as two AC/DC ILCs using a dual-droop (or integral dual-droop - see \cite{TPS_paper} for details) scheme. The outer loop equations are given by \eqref{eq:dd}, and the full model is \eqref{eq:dd} in conjunction with \eqref{eq:gfl_ilc}:
\begin{subequations}\label{eq:dd}
\begin{align}
    \dot \xi_1 &= K_{v1}\tilde{V}_{dc} - K_{\omega1}\omega_1 \\
    \dot \xi_2 &= K_{v2}\tilde{V}_{dc} - K_{\omega2}\omega_2 \label{eq:dd2_1}\\
    P_{ref,1} &= K_{v1}\tilde{V}_{dc} - K_{\omega1}\omega_1 + K_{i1} \xi_1\\
    P_{ref,2} &= K_{v1}\tilde{V}_{dc} - K_{\omega2}\omega_2 + K_{i2} \xi_2\label{eq:dd2_2}
\end{align}
\end{subequations}

\jdw{where $K_{v1}$, $K_{v2}$, $K_{\omega1}$, and $K_{\omega2}$ are the dual droop gains and $K_{i1}$, $K_{i2}$ are the integral gains. Note that it is possible to have different gains in the integrator (e.g. \eqref{eq:dd2_1}) and the power equation (e.g. \eqref{eq:dd2_2}), but for simplicity we keep them the same. } One advantage of this control scheme, in comparison to Dual Frequency Droop schemes, is that with this control the DC bus could be used to connect a DC grid and achieve power sharing in a hybrid AC/DC setting as in \cite{TPS_paper}. 
However, it has been suggested that introducing \jdw{additional coupling between the DC and AC sides} of the ILC may affect stability \cite{yoo2019}.

\subsection{AC Grid-forming controllers}
\jdw{In this case, the controller sets the reference frequencies $\omega_{ref,1}$ and $\omega_{ref,2}$ according to the outer loop design which we discuss in the this section. The equations which set the reference frequencies $\omega_{ref,1}$ and $\omega_{ref,2}$ are used in conjunction with the ILC model \eqref{eq:gfm_ilc} to form the reduced-order dynamical model of the AC/AC ILC. }

\subsubsection{Matching Control} \label{sect:dmc}
The "matching" control concept was proposed in \cite{jouini2016,arghir2018} for grid-forming inverters, with the idea of emulating a synchronous machine. However, it was also extended to ILCs in hybrid AC/DC networks \cite{TCST_paper}. The control laws are simple and \jdw{relate} the DC voltage to the AC frequencies. Since both AC frequencies are proportional to the DC voltage, they are also proportional to each other as desired for power sharing:
\begin{subequations}\label{eq:mc}
\begin{align}
    \omega_{ref,1} &= m_1\tilde{V}_{dc} \label{eq:mc1}\\
    \omega_{ref,2} &= m_2\tilde{V}_{dc}
\end{align}
\end{subequations}
and the full model (which we will call \emph{Matching} in the analysis) is given by \eqref{eq:gfm_ilc},\eqref{eq:mc}.

\subsubsection{Grid-forming Frequency Droop Control} \label{sect:gffd}
This scheme is based on the traditional frequency droop controller for inverters in AC microgrids, which is well known, e.g. \cite{chandorkar1993}:
\begin{equation}\label{eq:gfd_1}
    \omega_{ref,1} = - m_{p1}p^x_1
\end{equation}
where $m_{p1}$ is the droop gain which could be set to $1/K_{\omega1}$ for consistency (\jdwrev{i.e.,} the same power-frequency slope) with the GFL ILC controllers, $\omega_{ref,1}$ is the reference frequency deviation for the GFM converter, and $p^x_1$ is the power (deviation from a reference power) from the converter into the bus to which it is connected.

As stated, our required objective is that all the controllers must normalize the AC frequencies between sub-grids. Hence, to achieve this we consider the addition of an integral term (much like secondary control in an AC grid) $P_{eq}$. Furthermore, due to a requirement to regulate the DC voltage, an extra power term $P_{dc}$ is added to \eqref{eq:gfd_1} as in \cite{bala2009}:
\begin{subequations}\label{eq:gfd_2}
\begin{align}
    \dot \zeta &= \tilde{V}_{dc} \\
    P_{dc} &= K_{pdc}\tilde{V}_{dc} + K_{idc} \zeta \\
    \dot{P}_{eq} &= \omega_{ref,1} - \omega_{ref,2} \\
    \tau_1 \dot p^x_{f1} &= -p^x_{f1} + p^x_1 \\
    \tau_2 \dot p^x_{f2} &= -p^x_{f2} + p^x_2 \\
    \omega_{ref,1} &= - m_{p1}(p^x_{f1} + \kappa_{s1}{P_{dc}- K_{i1}P_{eq}} )\\
    \omega_{ref,2} &= - m_{p2}(p^x_{f2} + \kappa_{s2}{P_{dc}+ K_{i2}P_{eq}} )
\end{align}
\end{subequations}
where $\tau_1, \tau_2$ are the time constants of the power filters, \jdjdw{$\kappa_{s1}$ and $\kappa_{s2}$ are constants which allocate the sharing of the DC regulation and frequency-equalization terms (set to 0.5 in all numerical simulations)}, and $p^x_{f1}, p^x_{f2}$ are the filtered power \jdjdw{measurements} used in the control law \cite{chandorkar1993}.

\subsubsection{Grid-forming dual droop} \label{sect:gfdd}
The controller \eqref{eq:mc} varies the AC frequency based on the DC voltage, while \eqref{eq:gfd_2} varies the AC frequency of the converter based on power. By combining these concepts, we introduce what is essentially a grid-forming version of \eqref{eq:dd}; in fact, the form of new control law can be obtained by rearrangement \footnote{\jd{However, note that \eqref{eq:gfdd} is used with \eqref{eq:gfm_ilc} while \eqref{eq:dd} is used with \eqref{eq:gfl_ilc}.}}
\begin{subequations}\label{eq:gfdd}
\begin{align}
    \dot \xi_1 &= K_{v1}\tilde{V}_{dc} - K_{\omega1}\omega_{ref,1} \label{eq:gfdd1}\\
    \dot \xi_2 &= K_{v2}\tilde{V}_{dc} - K_{\omega2}\omega_{ref,2} \\
    \tau_1 \dot p^x_{f1} &= -p^x_{f1} + p^x_1 \\
    \tau_2 \dot p^x_{f2} &= -p^x_{f2} + p^x_2 \\
    \omega_{ref,1} &= m_{p1}(-p^x_{f1} + K_{v1}\tilde{V}_{dc}  + K_{i1} \xi_1) \label{eq:gfdd5}\\
    \omega_{ref,2} &= m_{p2}(-p^x_{f2} + K_{v2}\tilde{V}_{dc}  + K_{i2} \xi_2)
\end{align}
\end{subequations}
where $m_{p1}$ and $m_{p2}$ being the frequency-droop gains\footnote{\jd{\jdwjdw{These should not be equal} to $1/K_{\omega i}$, otherwise substituting \eqref{eq:gfdd5} into \eqref{eq:gfdd1} would result in the cancellation of the DC voltage term, \jdwrev{i.e.,} there would be no DC voltage regulation.}}. The full model is therefore given by \eqref{eq:gfm_ilc},\eqref{eq:gfdd}. An interesting observation is that this control bears striking similarity to the linearization of hybrid angle control \cite[eq. (24)]{hac2022}, if $K_{i1}, K_{i2}$ are set to zero. Furthermore, if the $-m_{p1}p^x_{f1}$ and $-m_{p2}p^x_{f2}$ terms \jd{were also} removed, this would recover Matching control \eqref{eq:mc}.

\section{Analysis}\label{analysis}

\subsection{Passivity analysis}\label{passanalysis}

For the passivity analysis, the models are linearized around the equilibrium point (in practice, $V_{dc}$ takes its nominal value at steady-state) with parameter values as in Table \ref{tab:csparams}. Since passivity requires the positive realness of the ILC transfer function $G(s)$, the eigenvalues of the Hermitian part of the transfer matrix $G(j\omega)+G^\ast(j\omega)$ at all frequencies must be positive. Fig. \ref{fig:comp2} shows the eigenvalues of the Hermitian part of the transfer matrix $G(j\omega)+G^\ast(j\omega)$ over frequency. The passivity property was also verified via solving an appropriate feasibility linear matrix inequality (\jdwrev{i.e.,} the KYP Lemma \cite{khalil1991}).


\begin{figure}[ht]
    \centering
    \includegraphics[width=0.47\textwidth]{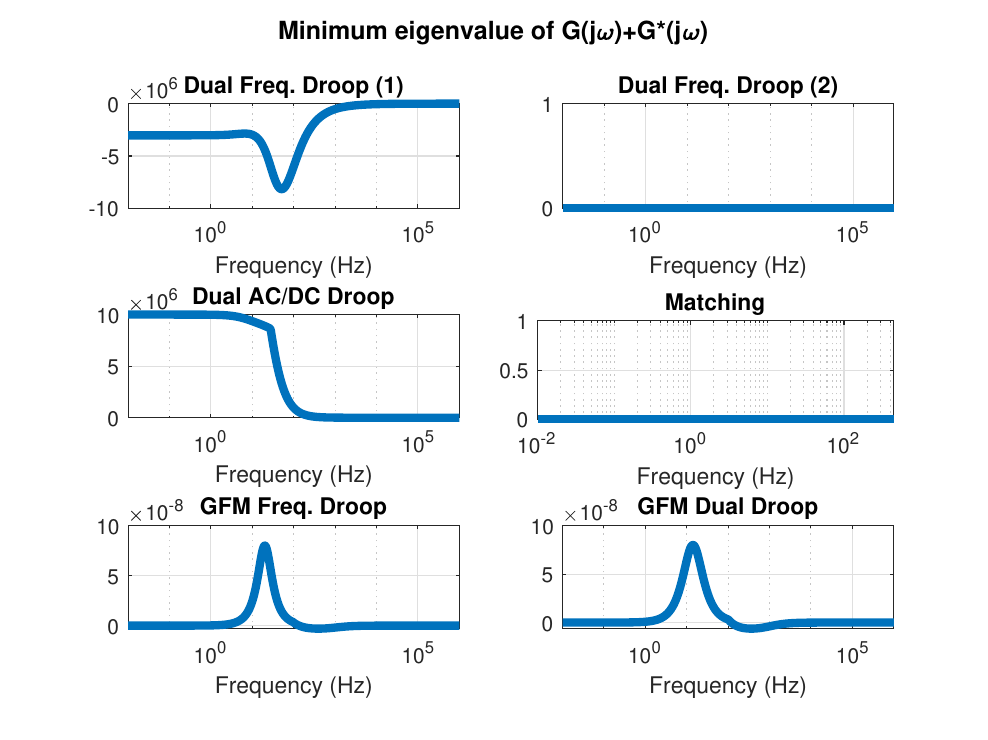}
    \vspace{-3mm}
    \caption{Passivity test at each frequency}
    \label{fig:comp2}
\end{figure}
There are several designs satisfying the passivity condition in Assumption \ref{ilcpass}, with others very close to \jd{satisfying the passivity condition} as well (and the grid-forming droop designs can also be passivated by \jd{adjusting} the parameters in Table \ref{tab:csparams}). \jdwrev{\jdwjdw{However,} designs in which one converter (e.g. the second converter as in \eqref{eq:dfd_2}) only regulates DC voltage cannot be passivated. This \lir{can be seen} as follows:}

\jdwrev{a) the change in frequency is detected by the first converter as in \eqref{eq:dfd_1}, which adjusts $p_1^x$, \ilrr{with the corresponding transfer function given by}:}
\begin{equation}
    \jdwrev{T_{(-\omega_2) \to p_1^x} = -(K_{w2} + \frac{K_i}{s})(\frac{1}{\tau s + 1})}
\end{equation}

\jdwrev{b) the change in power $p_1^x$ causes the DC voltage to deviate, \ilrr{with the \lir{corresponding} transfer function given by}:}
\begin{equation}
    \jdwrev{T_{p_1^x \to V_{dc}} = -\frac{1}{\frac{C_{dc}}{V_{dc}^*}s + K_{dc}}}
\end{equation}

\jdwrev{c) the second converter responds to that DC voltage deviation via closed-loop PI control \eqref{eq:dfd_2},  i.e.:}
\begin{equation}
    \jdwrev{T_{V_{dc} \to p_2^x} =\frac{K_{pdc} + \frac{K_{idc}}{s}}{1 + (K_{pdc} + \frac{K_{idc}}{s})(\frac{C_{dc}}{V_{dc}^*}s + K_{dc})}}
\end{equation}

\jdwrev{Hence the transfer function $T_{(-\omega_2) \to p_2^x}$ from frequency to power at the second converter is:}
\begin{equation}
    \jdwrev{\frac{(sK_{w2}+K_i)(sK_{pdc}+K_{idc})}{s(\tau s + 1)(\frac{C_{dc}}{V_{dc}^*}s+K_{dc})(s + (sK_{pdc} + K_{idc})(\frac{C_{dc}}{V_{dc}^*}s + K_{dc}))}}
\end{equation}

\jdwrev{Since the degree of the denominator exceeds that of the numerator by at least two (and this is the case for any choice of linear causal controller in $T_{(-\omega_2) \to p_1^x}$ or $T_{V_{dc} \to p_2^x}$), this results in a non-passive diagonal term on the MIMO transfer function. Since passivity requires $G(j\omega) + G^*(j\omega) \geq 0$, which is not possible if a diagonal term is negative, these schemes (unlike many other schemes which share DC voltage regulation tasks) cannot be made passive.}

\subsection{Case studies}\label{casestudy}

\subsubsection{Case study 1}

We consider a simple two-microgrid test network as shown in Fig. \ref{fig:network}, \jwrev{implemented in MATLAB/Simulink. In this case, synchronous generator dynamics govern the frequency of each of the two MGs. In this first case study, we simulate the nonlinear equations, while the second case study is more realistic and uses highly detailed models.} The network parameters are shown Fig. \ref{fig:network}, and also in Table \ref{tab:csparams} along with the converter and control parameters. For simplicity, the two converters which make up the AC-AC ILC have identical parameters, and the subscript $1$ and $2$ will be 
\icl{omitted}
hereafter.
\begin{figure}[ht]
    \centering
    \includegraphics[width=0.45\textwidth]{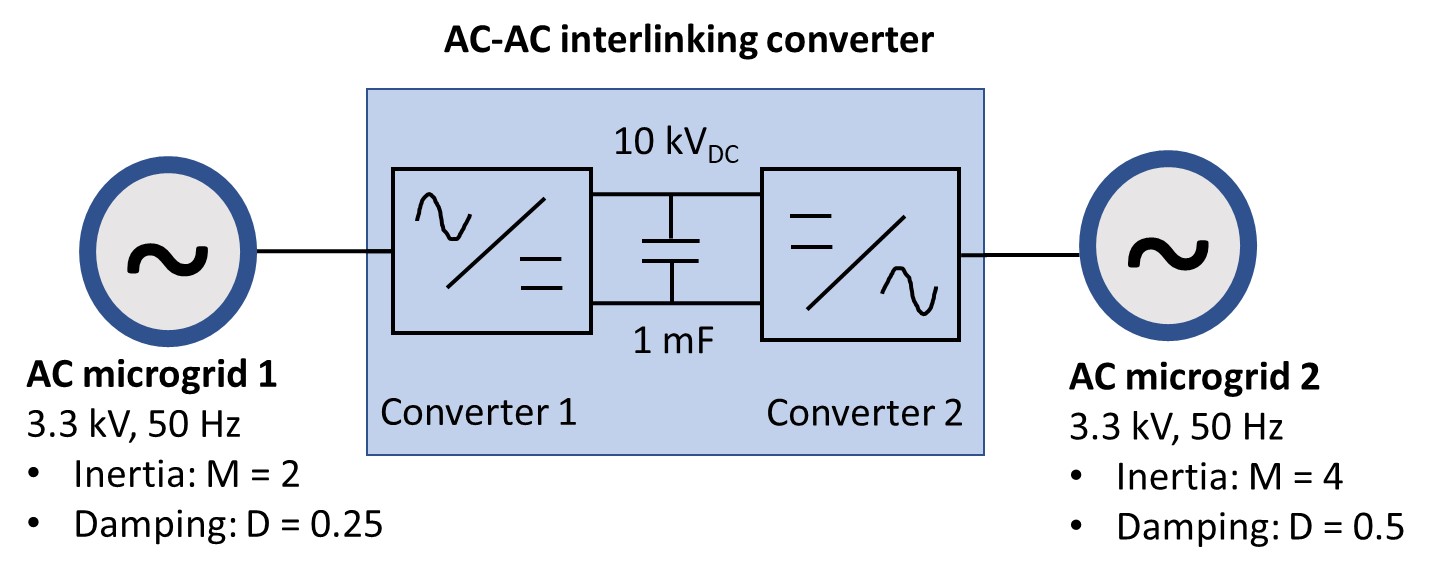}
    \vspace{-3mm}
    \caption{Network in case study 1}
    \label{fig:network}
\end{figure}

\begin{table}[ht]
	\centering
	\caption{Case study default parameters.}
		\begin{tabular}{c|c|c}
		Category & Parameter & Value \\
		\hline
		& DC voltage & $V_{dc,ref}$ = 10 kV \\
		Grid parameters & DC support & $K_{dc} = 1$ \\
		& AC voltage & $V_{ac}$ = 3.3 kV \\
		\hline	
		& DC capacitance & $C_{dc} = 1 mF$ \\
		ILC parameters & AC inductance & $L = 1 mH$ \\
		& ILC time constant & $\tau_1 = \tau_2 = 0.05 s$ \\
		\hline
		& Frequency droop & $K_{\omega 1} = K_{\omega 2} = 2.5 \times 10^7$ \\
		& Voltage droop & $K_{v1} = K_{v2} = 2.5 \times 10^4$ \\
		& AC/DC integral gain & $K_{i1} = K_{i2} = 10$ \\
		Control & Frequency integral gain & $K_{i1} = K_{i2} = 10$ \\
		Parameters & Matching gain & $m_1 = m_2 = 10^{-3}$ \\
            & Frequency droop gain & $m_{p1} = m_{p2} = 5 \times 10^{-8}$ \\
		& DC proportional gain & $K_{pdc} = K_{v1}$ \\
		& DC integral gain & $K_{idc} = 10 K_{pdc}$ \\
		\end{tabular}
    \vspace{-3mm}
	\label{tab:csparams}
\end{table}
The model is non-linear and more detailed than the analytical model used for passivity assessment. It also includes turbine-governor dynamics of the synchronous \jdwjdw{machines}, and we consider various load changes which can push the system away from the analyzed equilibrium point. Hence, it is possible for the system to be unstable.

An interesting point is that the direction of the initial \jd{(transient)} DC voltage response is topology-dependent for schemes that regulate the DC voltage with a single VSC. This is easily understood by taking the example of Dual Frequency Droop. Consider the control law $P_{ref,1} = -K_{\omega1}\omega_1 + K_{\omega2}\omega_2$ \eqref{eq:dfd_0}. The initial DC-AC power (dictated entirely by VSC 1) is in the opposite direction depending on which AC sub-grid the disturbance appears. Hence, the switching \icl{on} of a load at AC subgrid 1 would reduce $\omega_1$, thus increasing $P_{ref,1}$, leading to a decrease in DC voltage. However, the same load switched \icl{on} at AC subgrid 2 would reduce $\omega_2$, decreasing $P_{ref,1}$ initially, thus leading to a short-term \emph{\jdw{increase}} in the DC voltage. This is undesirable, as any source connected to the DC bus (e.g. using DC voltage droop control) would contribute \jd{to the power imbalance rather than compensating for it as desired. It also} precludes the use of the DC capacitance for inertial support. This is actually the fundamental contradiction between DC voltage control and inertial emulation recently discussed in \cite{rokrok2022}.

Given the default parameters in Table \ref{tab:csparams}, we adjust several important parameters and find the stability boundaries. \jwrev{Too little DC support ($K_{dc})$, too large a time constant or droop gains, or too high inductances can destabilize some of the models and the minimum (or maximum) value is shown in Table \ref{tab:csresults1}. \lir{For ease of} comprehension, \ilr{we use \textbf{bold} 
in the table for} the best stability regions attained for each parameter. } }

\begin{table}[ht]
\scriptsize
	\centering
	\caption{Case study 1 results.}
		\begin{tabular}{c|c|c|c|c}
					Scheme & Min $K_{dc}$ & Max $\tau$ & Max gain & Min $L$\tablefootnote{This is relevant to GFM schemes only. If the GFM VSC is too close electrically to the SG, oscillations may result.} \\
		    \hline	
		    Dual Freq. Droop (1) & \textbf{0.00} & 0.07 & $K_\omega = 1 \times10^9$ & N/A\\
		    Dual Freq. Droop (2) & \textbf{0.00} & 0.08 & \textbf{Any reasonable}  & N/A\\
		    Dual AC/DC Droop & \textbf{0.00} & 0.08 & \textbf{Any reasonable}  & N/A\\
		    Matching  & 0.06 & $\mathbf{>5}$ & $m = 0.03$ & 0.70 mH\\
		    GFM Freq. Droop & 0.20 & 0.09 & $m_p = 1 \times 10^{-6}$ & 0.30 mH\\
		    GFM Dual Droop & 0.08 & 0.50 & $m_p = 6 \times 10^{-6}$ & 0.05 mH \\
		    \hline
		    Dual Droop+Matching & \textbf{0.00} & $\mathbf{>5}$ & \textbf{Any reasonable} & \textbf{0.01 mH} \\
		    GFL+GFM Dual Droop & \textbf{0.00} & $\mathbf{>5}$ & $m_p = 9 \times 10^{-6}$ & \textbf{0.01 mH} \\
		\end{tabular}
	\label{tab:csresults1}
\end{table}

Note that the dual AC grid-forming schemes require DC support (\jdwrev{i.e.,} $K_{dc} > 0$), while the AC grid-following schemes do not require this, \jdwrev{i.e.,} they are stable for $K_{dc} \to 0$. However, as the time constant $\tau$ is increased, the grid-forming schemes perform better. This is consistent with various studies showing improve stability via AC grid-forming schemes when appropriate DC support is present, \jdw{e.g. \cite{zuo2021}}.

\icl{{\em Partially grid-forming ILC.}}
In view of this, \jdwjdw{it is worth considering having one inverter in AC grid-forming mode, while the other remains grid-following (\jdwrev{i.e.,} a "partially grid-forming ILC")}. This allows the grid-forming inverter to provide support to the weaker AC sub-grid, while the grid-following inverter connects to the stiffer grid as is advantageous \cite{zhao2022}. Numerical analysis shows that the \jd{grid-following \emph{Dual AC/DC Droop} with the grid-forming \emph{Matching} control scheme \jd{\eqref{eq:mix_ilc}, \eqref{eq:dd2_1}, \eqref{eq:dd2_2}, \eqref{eq:mc1}} is an effective choice}. \jd{This} configuration has a larger stability region in the case study (second-last row of Table \ref{tab:csresults1}) than the only-GFL or only-GFM alternatives, and is able to provide inertial \jdwjdw{support}. In addition, \jd{this scheme satisfies Assumption \ref{ilcpass}}\footnote{\jd{Passivity was verified via solving an appropriate LMI, and equilibrium-state observability is shown in Appendix C.}} with the default parameters in Table \ref{tab:csparams}. 

\subsubsection{Case study 2}
\begin{figure}[ht]
    \centering
    \includegraphics[width=0.5\textwidth]{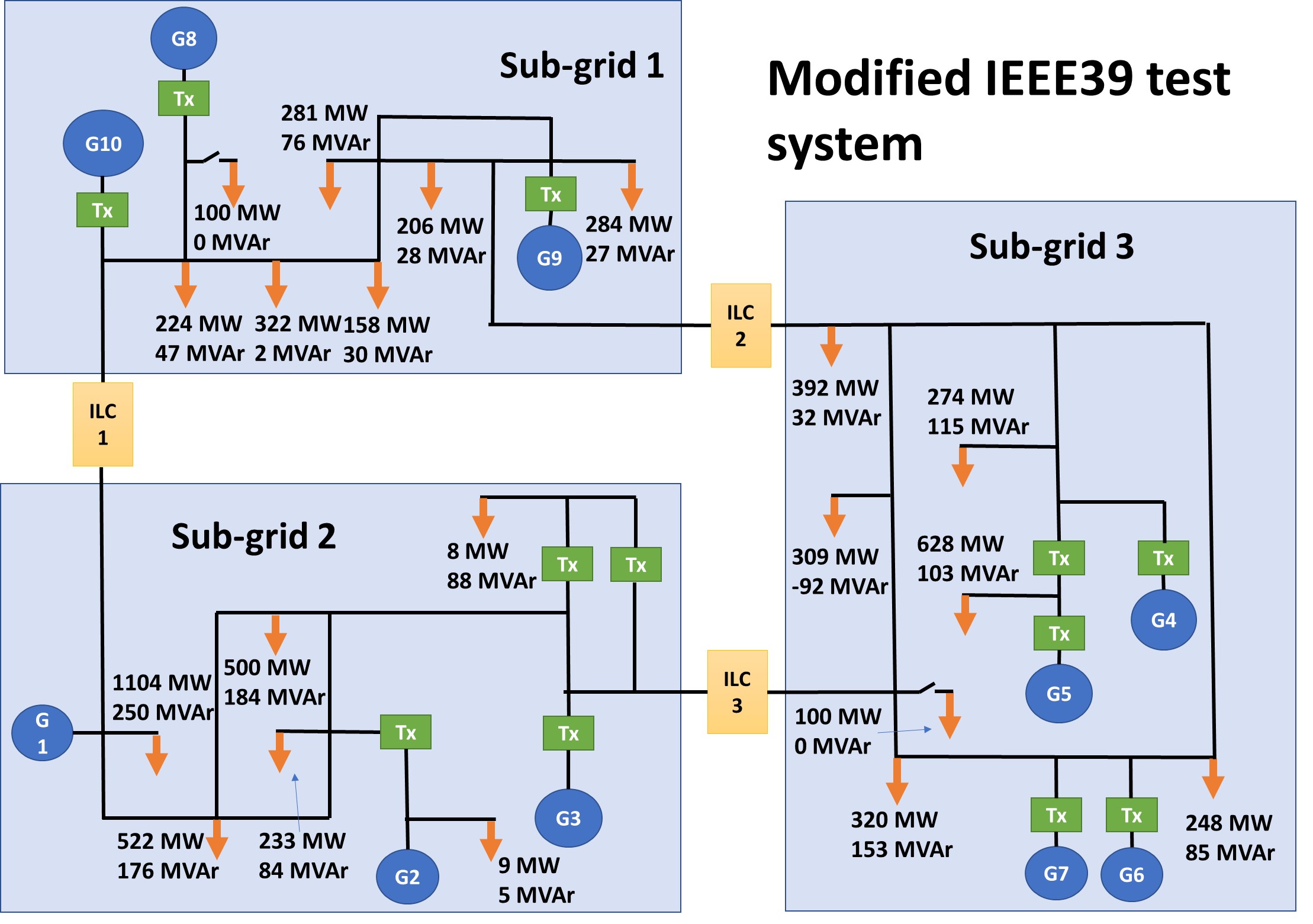}
    \caption{\jwrev{Modified IEEE39 test system}}
    \label{fig:ieee39}
\end{figure}

We apply the proposed ILC control algorithms to the well-known IEEE New England 39-bus system \cite{newengland39}. The model is detailed (\jwrev{considerably more detailed than the analytical model presented earlier}) and includes high-order models of the generators, turbine-governors, exciters, transformers, and lines; \jwrev{this model is implemented in MATLAB/Simscape Electrical. Minor modifications are made to illustrate the use of AC-AC ILCs; the network is transformed into a three MG / three ILC system as shown in Fig. \ref{fig:ieee39}.}

\begin{figure}[ht]
    \centering
    \includegraphics[width=0.5\textwidth]{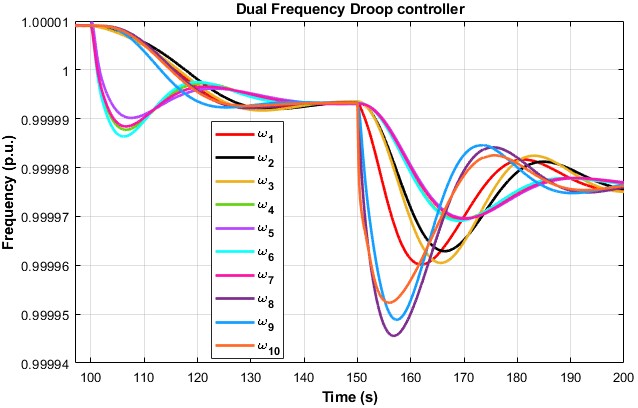}
      \vspace{-2mm}
    \caption{\jwrev{Dual Frequency Droop (1) controller in the IEEE 39 bus system: \jdw{\eqref{eq:gfl_ilc}, \eqref{eq:dfd_1}, \eqref{eq:dfd_2}}}}
    \label{fig:dfd}
\end{figure}

\begin{figure}[ht]
\vspace{-2mm}
    \centering
    \includegraphics[width=0.5\textwidth]{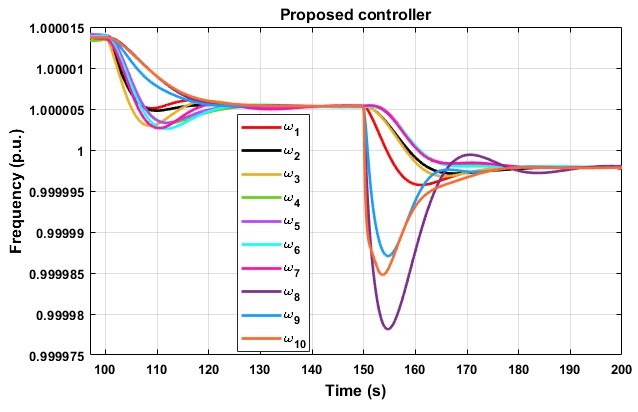}
    \vspace{-2mm}
    \caption{\jwrev{Dual AC/DC Droop + Matching controller in the IEEE 39 bus system: \jdw{\eqref{eq:mix_ilc}, \eqref{eq:dd2_1}, \eqref{eq:dd2_2}, \eqref{eq:mc1}}}}
    \label{fig:gfdd}
\end{figure}

This results in three AC subsystems connected by the three ILCs, each subsystem containing generators 1-3, 4-7 and 8-10 respectively.  The ILC is required to equalize the AC frequencies of both subsystems and thus achieve proportional power sharing.
\jjddww{
\icc{This raises the possibility of oscillations depending on the control policy of the AC-AC
ILC, although} the system remains stable for reasonable control schemes and parameters. \icc{Differences in damping and response speed can, however,}} be noted, as we will \jd{illustrate}. We compare the dual frequency droop scheme to two alternatives in Figs. \ref{fig:dfd}-\ref{fig:gfdd}, using the same parameters as in Table \ref{tab:csparams} except for the following: the DC voltage is now 500 kV (HVDC); the DC capacitance is reduced to 0.1 mF; and the factor $K_{dc}$ is reduced by 100 in view of a much higher DC voltage.

\begin{figure*}[!ht]
    \centering
    \includegraphics[width=0.95\textwidth]{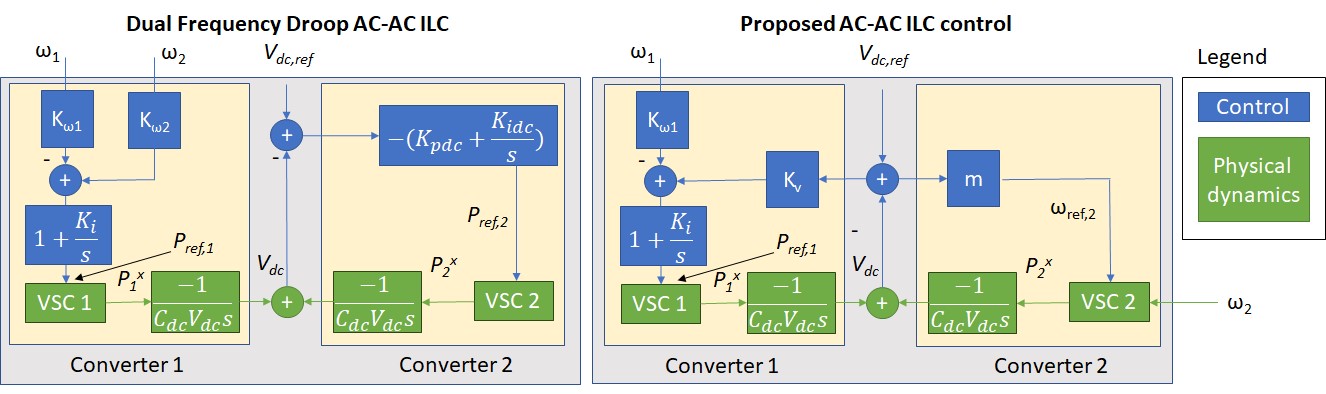}
    \caption{\jwrev{Controllers compared in the second case study}}
    \label{fig:controllers}
\end{figure*}

In this \jwrev{case study}, we simply compare the conventional Dual Frequency Droop (1) controller to our proposal of Dual AC/DC Droop + Matching control
\icl{mentioned in the previous section}. Figs. \ref{fig:dfd} - \ref{fig:gfdd} show the performance under these conditions. Loads of 50MW are added at bus 24 at t = 150s, and similarly at bus 25 at t = 200s. \jwrev{The controllers are illustrated in Fig. \ref{fig:controllers}.}

The control gains were tuned for performance. Compared to the parameters in Table \ref{tab:csparams}, the AC frequency droop gain was increased by a factor of 1000 in view of a higher-rated system. 
In the case of Dual AC/DC Droop + Matching control, the DC voltage droop was reduced by a factor of 10 in view of a higher DC voltage, and $m = 0.01 \frac{K_{vdc}}{K_\omega}$ was used for the matching control. \jwrev{For each ILC, when partially grid-forming control is used, the grid-following converter is always connected to the stronger subsystem, and the grid-forming side is connected to the weaker grid in Fig. \ref{fig:ieee39}; \lir{i.e.,} ILC 1 is grid-forming on the "Sub-grid 1" side, ILC 2 is grid-forming on the "Sub-grid 3" side, and ILC 3 is also grid-forming on the "Sub-grid 3" side. }
\vspace{-0mm}

\jdwrev{Figs. \ref{fig:dfd}-\ref{fig:gfdd} show the simulation results in this case, showing the per-unit frequency of each of the ten synchronous generators of the IEEE 39 model (per Fig. \ref{fig:ieee39}, generators 1-3 with frequencies $\omega_1, \omega_2, \omega_3$ are in sub-grid 2, generators 4-7 are in sub-grid 3, and generators 8-10 form sub-grid 1).} The partially grid-forming (Dual AC/DC Droop + Matching) ILC responds better to disturbances. The Dual Frequency Droop controller clearly provides less damping. allows the frequency nadir to drop around 30-40\% further upon a load change, and takes considerably longer to reach steady-state. The Dual AC/DC Droop + Matching controller also has a considerably larger stability region, \jdwrev{i.e.,} it was possible to operate stably with higher droop gains (both AC frequency and DC voltage). This improved performance is for a number of reasons: a) as demonstrated by the passivity analysis, better power/frequency damping is provided; b) by sharing DC voltage regulation and using GFM control, the DC-side inertia and support ($K_{dc}$) can be used to help stabilize the AC system. 

Hence, the idea of "partially grid-forming ILCs" to use one inverter in grid-forming mode to strengthen the weaker grid while the other connects to the stronger grid in grid-following mode is of benefit in this \jwrev{case study}. \jdw{This is consistent with recent findings showing that a significant proportion of ILCs in AC GFM mode can improve stability \cite{soler2023}.} 




\subsubsection{Summary}
\jdwjdw{The following points are notable:}
\begin{itemize}
    \item \jdwjdw{From a stability perspective, it is better if DC voltage regulation is shared between the two VSCs.}
    \item \jd{Our analysis also addresses the question \cite{yoo2017} of whether the control of the AC/AC ILC should couple frequency control and DC voltage regulation, a feature present in some of the schemes in Table \ref{tab:litacac}. In our studies, schemes that use the DC voltage generally have better stability properties, and \jdwjdw{this also enables} the provision of inertia from the ILC. This coincides with the success of recent schemes for GFM converters which couple frequency and DC voltage dynamics, e.g. \cite{hac2022}.}
    \item Dual AC grid-forming behavior requires strong support on the shared DC bus. Hence, a good alternative is partial grid-forming, \jdwrev{i.e.,} AC grid-forming behavior only on one side (\ilrr{the} weaker AC sub-grid).
\end{itemize}



\section{Conclusion}\label{concl}

\jdwjdw{We have considered the problem of controlling multi-grids to ensure stability and power sharing. A passivity framework was proposed for their stabilization, resulting in a decentralized stability condition \jd{which}, when satisfied, guarantees the stability of the \il{multi-MG grid}.} 
After reviewing AC/AC ILC control methods in the literature, we propose a controller which is verified by simulation results and theoretical analysis. \jd{This controller allows AC grid-forming capability on one side, which should be connected to the weaker AC sub-grid, while still satisfying the passivity condition for stability.} Further work \jd{includes} adapting the passivity framework to include DC microgrids, and considering reactive power / voltage control as well.


\section*{Appendix}
\jd{We provide the proof of our main result in Appendix A and show that the proposed ILC controller satisfies Assumption 2b in Appendix C.}
\subsection{Proof of Theorem \ref{stabtheorem}}
\jd{We use $\mathscr{W}(q) = \sum_{j=1}^{|N|}W_j(x_j) + \ill{\sum_{l=1}^{|Z|}\tilde{W}_{l}(z_{l})}$, as a candidate Lyapunov function where $q=(x_1,x_2, \dots, x_{|N|},z_1,z_2, \dots, z_{|E|})$ \ill{and $W_j$, $\tilde W_l$ are the storage functions associated with Assumptions 2a, 3a.}
\jd{
} \ill{We consider a \icc{neighbourhood 
of $q$} such that the properties in \icc{Assumptions \ref{mgpass}a, \ref{ilcpass}a} hold. \icc{Differentiating w.r.t. time we have}} 
\begin{equation}
    \dot{\mathscr{W}} \leq  \sum_{j=1}^{|N|}(\omega_j-\omega_j^*)\phi_j(\omega_j-\omega_j^*)
\end{equation}
where $\phi_j$ is a positive definite function for all $j \in N$. $\mathscr{W}(q)$ is therefore non-increasing with time, and from \ill{Assumption~1} 
$q^*$ is a strict local minimum of $\mathscr{W}$.
}

\jd{The Theorem then follows by applying LaSalle's theorem~\cite{khalil1991} with $\mathscr{W}(q)$ as the Lyapunov-like \icl{function.}} 
\jd{
\ill{Consider the level set $\mathcal{T}=\{q:\mathscr{W}(q)-\mathscr{W}(q^\ast) \leq \epsilon\}$, with $\epsilon>0$ sufficiently \icc{small such that 
$\mathcal{T}$ is compact and~positively~invariant, and the properties in Assumptions \ref{mgpass}, \ref{ilcpass} hold.}
LaSalle's} theorem states that all solutions with initial conditions within $\mathcal{T}$ converge to the largest \ill{invariant set within $\mathcal{T}$ for} which $\dot{\mathscr{W}}=0$. \ill{We denote this invariant set as $\mathcal{S}$.}}

\jd{$\dot{\mathscr{W}}=0$ implies $\omega_j=\omega_j^*$ for all $j \in N$, and via \ill{Assumption 2c} 
further implies that the inputs to the MG systems are also constant equilibrium values, \jdwrev{i.e.,} $p_j=p_j^*$ for all $j \in N$. 
Since the MG dynamics are asymptotically stable for constant input \ill{(Assumption 2b)}, \ill{for $\epsilon$ sufficiently small each $x_j$ \icc{in $\mathcal{S}$}} converges to $x^*_j$ when $p_j=p_j^*$. Since $\mathscr{W}(q)$ has a strict local minimum at $q^*$, and is constant w.r.t. \ill{time in $\mathcal{S}$, we must have  $x_j = x^*_j$ in $\mathcal{S}$.} 
Finally, $p_j=p_j^*$ and $\omega_j=\omega_j^*$ imply constant inputs to, and constant outputs from, the ILC systems. Since these systems are equilibrium-state observable \ill{(Assumption 3b)}, this implies also that $z_l = z_l^*$ from all $l$. Hence, the only invariant set within $\mathcal{T}$ for which $\dot{\mathscr{W}}=0$ is the equilibrium point $q^*$, and \ill{therefore} this equilibrium point is locally asymptotically~stable. }

\subsection{\jwrev{Observability definitions}}
\begin{definition}[Equilibrium-state observability]\label{def:eso}
\jd{The system \eqref{eq:ss} is \jjdw{locally equilibrium-state observable about an equilibrium point $(u^*,x^*)$} if \icl{there exists a neighbourhood $\Omega$ of $x^\ast$ such that $u(t) \equiv u^*$, $x(t)\in\Omega$ and $y(t) \equiv y^*$ for all $t$} implies $x(t) \equiv x^*$. Equivalently, no solution of $\dot{x} = f(x, u^*)$ can stay identically in \icl{$S = \{x \in \mathbb{R}^n : x\in\Omega, y = g(x, u^*) = y^*\}$}, other than the solution $x(t) \equiv x^*$. We will call this property \emph{equilibrium-state observability}.}
\end{definition}

\begin{remark}
    Equilibrium-state observability in Definition \ref{def:eso} is an adaptation of the more well-known zero-state observability\cite{khalil1991} for non-zero equilibria.
\end{remark}

\begin{definition}[Input observability \cite{boukhobza2007}]\label{def:iobsrv}
\jjdw{The system \eqref{eq:ss} is locally input observable about an equilibrium point $(u^*,x^*)$ if \icl{there exists a neighbourhood $\Omega$ of $x^\ast$ such that} $y(t) \equiv y^*$, \icl{$x(t)\in\Omega$ for all $t$ implies} $u(t) \equiv u^*$.
Equivalently, there is no trajectory of $\dot{x} = f(x, u)$ that stays in \icl{$S = \{x : x\in\Omega, y = g(x,u) = y^*\}$} without $u(t) \equiv u^*$. We will call this property \emph{input observability}.}
\end{definition} 

\subsection{Equilibrium-state observability}

\jd{The proposed partially grid-forming ILC controller \jjdw{Sections \ref{sectpartiallygfm}  Section \ref{casestudy}} has the dynamics \eqref{eq:mix_ilc}, \eqref{eq:dd2_1}, \eqref{eq:dd2_2}, \eqref{eq:mc1} which can be written as:
\begin{subequations}\label{eq:pgfm}
\begin{align}
    \dot \eta &= m\tilde{V}_{dc} - \omega_1 \\
    \dot \xi_2 &= K_{v2}\tilde{V}_{dc} - K_{\omega2}\omega_2 \\
    \tau_2 \dot p^x_2 & = - p^x_2 + K_{v1}\tilde{V}_{dc} - K_{\omega2}\omega_2 + K_{i2} \xi_2 \\
    C \dot{\tilde{V}}_{dc} & = - \frac{B\eta}{\tilde{V}_{dc} + V_{dc,ref}} - \frac{p^x_2}{\tilde{V}_{dc} + V_{dc,ref}} - K_{dc}\tilde{V}_{dc}
\end{align}
\end{subequations}}

\jd{We now aim to show that no solution of \eqref{eq:pgfm} can stay identically in \icl{$S = \{z \in \mathbb{R}^n : y = g(z, u^*) = y^*\}$,} other than the solution $z(t) \equiv z^*$, where $z = [\eta \mbox{  } \xi_2 \mbox{  } p^x_2 \mbox{  } \tilde{V}_{dc}]^T$.} \jd{$u = u^*$ and $y = y^*$ implies that inputs $-\omega_1 = u_1^*,\mbox{ } -\omega_2 = u_2^*$ and outputs $p^x_1 = B\eta = B\eta^*, p^x_2 = p^{x*}_2$ are constant within $S$. Any solution that stays identically in $S$ must therefore also have $\dot \eta = 0$ and $\dot p^x_2 = 0$. Substituting this into \eqref{eq:pgfm}:
\begin{subequations}\label{eq:pgfm2}
\begin{align}
    0 &= m\tilde{V}_{dc} - \omega^*_1 \label{eq:pgfm2a}\\
    \dot \xi_2 &= K_{v2}\tilde{V}_{dc} - K_{\omega2}\omega^*_2 \\
    0 & = - p^{x*}_2 + K_{v1}\tilde{V}_{dc} - K_{\omega2}\omega^*_2 + K_{i2} \xi_2 \label{eq:pgfm2c}\\
    C \dot{\tilde{V}}_{dc} & = - \frac{B\eta^*}{\tilde{V}_{dc} + V_{dc,ref}} - \frac{p^{x*}_2}{\tilde{V}_{dc} + V_{dc,ref}} - K_{dc}\tilde{V}_{dc}
\end{align}
\end{subequations}}

\jd{\eqref{eq:pgfm2a} shows that, within $S$, $m\tilde{V}_{dc} = \omega^*_1$ and therefore $\tilde{V}_{dc} = \tilde{V}^*_{dc}$ within $S$ since $m > 0$. Thus $\dot{\tilde{V}}_{dc} = 0$. Next, consider \eqref{eq:pgfm2c}. All terms of \eqref{eq:pgfm2c} are known to be constant within $S$ except $\xi_2$, hence $\xi_2$ must also be constant within $S$, \jdwrev{i.e.,} $\xi_2 = \xi^*_2$.} \jd{Hence, all the states of \eqref{eq:pgfm} are constant values within $S$, \jdwrev{i.e.,} $z(t) \equiv z^*$ within $S$. This completes the proof. }

\vfill


\end{document}